%% file: 0_main.tex
\documentclass[sigconf]{acmart}
\AtBeginDocument{%
  }

\setcopyright{acmlicensed}
\copyrightyear{2018}
\acmYear{2018}
\acmDOI{XXXXXXX.XXXXXXX}
\acmConference[Conference acronym 'XX]{Make sure to enter the correct
  conference title from your rights confirmation email}{June 03--05,
  2018}{Woodstock, NY}
\acmISBN{978-1-4503-XXXX-X/2018/06}

\usepackage{CJK}

\usepackage{paralist}
\usepackage{enumitem}
\usepackage{pifont}
\usepackage{makecell}
\usepackage{color}
\usepackage{xcolor}
\usepackage[ruled, linesnumbered]{algorithm2e}
\usepackage{amsthm}
\usepackage{multirow}
\usepackage{subfigure}  
\usepackage{subcaption}
\usepackage{float}
\usepackage{wrapfig}
\usepackage{subfloat}
\usepackage{caption}
\usepackage{adjustbox}
\usepackage{mathtools}  
\usepackage{mathrsfs}   
\usepackage{graphicx}
\newcommand{\Rmnum}[1]{\expandafter\@slowromancap\romannumeral #1@} 
\usepackage{arydshln} 
\usepackage{tabularx}
\usepackage{array}
\usepackage{tabularx} 
\usepackage{makecell}
\usepackage{booktabs}

\newlength{\equalcolwidth}

\begin{document}

\title{S$^2$GR: Stepwise Semantic-Guided Reasoning in Latent Space for Generative Recommendation}


\author{Zihao Guo$^*$}
\affiliation{%
  \institution{Kuaishou Technology}
  \city{Beijing}
  \country{China}
}
\email{guozihao05@kuaishou.com}

\author{Jian Wang$^*$}
\affiliation{%
  \institution{Kuaishou Technology}
  \city{Beijing}
  \country{China}
}
\email{wangjian29@kuaishou.com}

\author{Ruxin Zhou}
\affiliation{%
  \institution{Kuaishou Technology}
  \city{Beijing}
  \country{China}
}
\email{zhouruxin@kuaishou.com}

\author{Youhua Liu}
\affiliation{%
  \institution{Kuaishou Technology}
  \city{Beijing}
  \country{China}
}
\email{liuyouhua@kuaishou.com}

\author{Jiawei Guo}
\affiliation{%
  \institution{Kuaishou Technology}
  \city{Beijing}
  \country{China}
}
\email{guojiawei03@kuaishou.com}

\author{Jun Zhao}
\affiliation{%
  \institution{Kuaishou Technology}
  \city{Beijing}
  \country{China}
}
\email{zhaojun11@kuaishou.com}

\author{Xiaoxiao Xu}
\affiliation{%
  \institution{Kuaishou Technology}
  \city{Beijing}
  \country{China}
}
\email{xuxiaoxiao05@kuaishou.com}

\author{Yongqi Liu}
\affiliation{%
  \institution{Kuaishou Technology}
  \city{Beijing}
  \country{China}
}
\email{liuyongqi@kuaishou.com}

\author{Kaiqiao Zhan}
\affiliation{%
  \institution{Kuaishou Technology}
  \city{Beijing}
  \country{China}
}
\email{zhankaiqiao@kuaishou.com}

\begin{abstract}
Generative Recommendation (GR) has emerged as a transformative paradigm with its end-to-end generation advantages. However, existing GR methods primarily focus on direct semantic ID (SID) generation from interaction sequences, failing to activate deeper reasoning capabilities analogous to those in large language models and thus limiting performance potential. We identify two critical limitations in current reasoning-enhanced GR approaches: (1) Strict sequential separation between reasoning and generation steps creates imbalanced computational focus across hierarchical SID codes, degrading quality for SID codes; (2) Generated reasoning vectors lack interpretable semantics, while reasoning paths suffer from unverifiable supervision. In this paper, we propose \underline{s}tepwise \underline{s}emantic-\underline{g}uided \underline{r}easoning in latent space (S$^2$GR), a novel reasoning enhanced GR framework. First, we establish a robust semantic foundation via codebook optimization, integrating item co-occurrence relationship to capture behavioral patterns, and load balancing and uniformity objectives that maximize codebook utilization while reinforcing coarse-to-fine semantic hierarchies. Our core innovation introduces the stepwise reasoning mechanism inserting thinking tokens before each SID generation step, where each token explicitly represents coarse-grained semantics supervised via contrastive learning against ground-truth codebook cluster distributions ensuring physically grounded reasoning paths and balanced computational focus across all SID codes. Extensive experiments demonstrate the superiority of~S$^2$GR, and online A/B test confirms efficacy on large-scale industrial short video platform.

\end{abstract}




\maketitle

\def\thefootnote{*}\footnotetext{Equal contribution.}

\input{1_introduction}
\input{2_related_work}

\input{3_preliminary}

\input{4_method}


\input{5_experiment}

\input{6_conclusion}



\bibliographystyle{ACM-Reference-Format}
\bibliography{kdd26}

\input{7_appendix}

\end{document}

%% file: 1_introduction.tex
\section{INTRODUCTION}

Recommendation systems, due to the ability to predict user preferences and behaviors by analyzing historical interactions, are essential for enhancing user experiences across platforms such as e-commerce, video streaming, and news portals. Recently, generative recommendation (GR) has emerged as a transformative paradigm, which replaces traditional large item ID space with a set of semantic tokens for item representation. By leveraging the content representation capability of semantic tokens and intent understanding capacity of generative models, GR enables end-to-end generation of items consistent with users' latent preferences directly from a semantic ID (SID) space, which overcomes limitations like objective misalignment in traditional multi-stage cascaded pipeline. Despite significant advancements in recent work, most existing methods primarily concentrate on generating SIDs directly from user interaction sequences. This constrained computational depth fails to effectively activate the generative models' reasoning potential, which have demonstrated significant effectiveness in large language models (LLMs).

\begin{figure*}[t]
\centering
\includegraphics[width=1.0\textwidth]{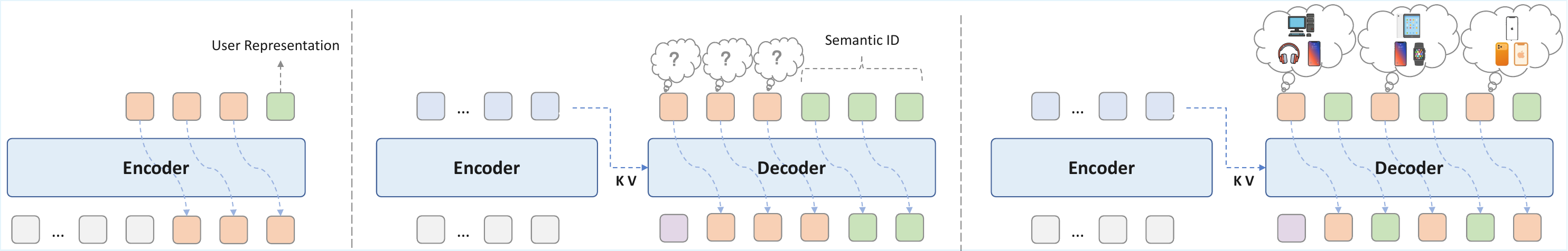} 
\caption{The illustrations of latent reasoning paradigms. The left and middle subfigures show the processes in sequential and generative recommendation, where multiple latent representations are autoregressively generated and refined before generating the final user representation or SID. The right subfigure presents our proposed stepwise reasoning paradigm for generative recommendation, wherein coarse-grained semantic categories are thinked before each SID code generation, serving as an transitional step toward the next SID code.}
\label{fig:scenario}
\end{figure*}

Recent efforts have explored latent reasoning paradigms in recommendation systems. ReaRec~\cite{tang2025think}, LARES~\cite{liu2025lares}, LCR-SER~\cite{shi2025bridging}, OnePiece~\cite{dai2025onepiece} have investigated implicit reasoning in sequential recommendation through autoregressive generation of multi-step latent status to refine user representations. For GR paradigm, STREAM-Rec~\cite{zhang2025slow} directly fills reasoning chain with similar item representations, while GPR~\cite{zhang2025gpr} synthesizes latent reasoning status from intent embeddings refined via diffusion. These methods collectively exhibit two potential limitations: (1) They strictly position the reasoning steps before any final output, forming a unidirectional chain from thinking block to generation block. However, the multi-code representation for SIDs constitutes multi-step generation, terminating the reasoning process before SID generation results in imbalanced computational distribution, failing to guarantee quality across all SID codes. Meanwhile, the potential semantic drift in early SID codes may pose a risk of cross-step error accumulation during autoregressive generation. (2) The generated reasoning status lack interpretable physical meaning, while the reasoning paths themselves suffer from the absence of explainable supervision signals. This dual deficiency fundamentally undermines the verifiable validity of reasoning path. Consequently, despite notable architectural innovations, the core challenge of eliciting reliably reasoning capabilities within GR remains unresolved.

To address these challenges, we conduct analysis from the perspective of inherent characteristics in GR. The items' SIDs derive from residual quantization, naturally possessing hierarchical semantics from coarse to fine granularity. The sequential generation of SID codes in GR inherently mirrors progressive recommendation decisions: initial high-level category selection (e.g., "electronics"), followed by progressive subclass refinement (e.g., "mobile phone"), culminating in specific item identification (e.g., "iPhone 17"). To adapt to this structural correspondence, as shown in Figure~\ref{fig:scenario}, we introduce thinking tokens interleaved with SID code generation to activate stepwise reasoning. Since SID generation inherently progresses through granularity levels, inspired by~\cite{li2025latent,bigverdi2025perception,chen2025think,zhang2025latent,qin2025chain}, we regard each thinking token before the SID codes as a transitional thinking module that reasonings relevant coarse-grained semantic clusters for the subsequent SID code generation step (e.g., representing several categories which are semantically close to "electronics" before first-code generation). Crucially, these thinking tokens acquire tangible physical meaning by design, they directly correspond to coarse-level semantic distributions. This enables stepwise supervision during training, where we constrain thinking tokens using ground-truth coarse SID semantics to ensure reasoning path reliability. Consequently, each SID code is generated after validated semantic thinking, significantly improving the final output quality and computational focus across all code positions.

Driven by above insights, we propose stepwise semantic-guided reasoning (S$^2$GR) in latent space to enhance GR. First of all, to ensure robust codebook quality that reinforces the coarse-to-fine semantic hierarchy, we align item semantic and user behaviours throught item co-occurrence relationships builded by historical user interactions, then optimize codebook distribution and utilization during quantization through load balancing and uniformity objectives. Crucially, the core of~S$^2$GR~lies in its stepwise latent reasoning mechanism with hierarchical semantic. We interleave thinking tokens before each SID code generation step, which are supervised via contrastive learning grounded in the coarse-grained clustering semantics of the codebook at the corresponding layer, thereby explicitly representing the categorical semantics relevant to subsequent SID prediction. Additionally, a additional lightweight decoder trained with in-batch negative sampling generates holistic user interest representations that provide auxiliary supervision for the initial thinking token, incorporating comprehensive item consideration at the generation onset, thereby mitigating semantic shifts in early codes. Our contributions are summarized as follows:
\begin{itemize}[leftmargin=0.3cm]
    \item We propose the stepwise latent reasoning that aligns with SID's hierarchical semantics and incorporates physically interpretable supervision to ensure reliable reasoning paths in GR.
    \item We enhance reasoning capability through comprehensive codebook optimization integrating collaborative signals, distribution uniformity, and load balancing.
    \item We validate our approach on public dataset and large-scale industrial dataset, demonstrating significant superiority over state-of-the-art baselines and achieving significant business gains in online A/B test on large-scale industrial short video platform.
\end{itemize}

%% file: 2_related_work.tex
\section{RELATED WORKS}

\subsection{Generative Recommendation}

Generative Recommendation (GR) represents a novel paradigm that departs from conventional multi-stage retrieval pipelines. By generating recommendations end-to-end within the semantic ID (SID) space, GR eliminates the need for intermediate retrieval stages and addresses objective misalignment through a holistic optimization of the final recommendation goals. TIGER~\cite{rajput2023recommender} pioneered this paradigm by compressing item representations from the vast original item ID space into a semantic space composed of ordered codewords through residual quantization, where item sequences represented by SIDs are then fed into generative models for training. Building upon this foundation, OneRec~\cite{deng2025onerec, zhou2025onerec, zhou2025onerec2} significantly enhanced the usability of GR via comprehensive improvements in model architecture and reinforcement learning. COBRA~\cite{yang2025sparse} further addressed information loss caused by the separation of SID quantization and sequence modeling stages by integrating traditional dense vector retrieval. GPR~\cite{zhang2025gpr} introduced targeted modeling for the more complex heterogeneity inherent in advertising recommendation scenarios, employing innovations like unified semantic representation, a dual-decoder architecture, and multi-stage joint training to balance efficiency and business alignment. However, most existing methods predominantly focus on direct SID generation without activating deeper reasoning capabilities akin to those in large language models (LLMs), which creates a bottleneck in the performance improvement of GR models.

\subsection{Reasoning in Recommender System}

With the advancement of LLMs, leveraging chain-of-thought (CoT) reasoning, which involves generating a sequence of tokens representing the thought process before producing the final answer, has opened promising pathways to enhance model reasoning capabilities~\cite{zhang2025landscape, zhang2025survey}. Inspired by similar concepts, how to endow recommendation systems with powerful reasoning abilities has garnered widespread attention~\cite{peng2025survey, zhang2025survey1, li2025survey}. 

Existing explorations fall into two lines: explicit reasoning and implicit reasoning. The explicit reasoning aims to reutilize the inherent reasoning capabilities of LLM by feeding candidate items as textual descriptions~\cite{zhang2025llmtreerec, bao2023tallrec, bao2025bi, chen2024softmax} or ID tokens~\cite{liu2025onerec,he2025plum} as input. This allows the model to perceive information based on its general world knowledge, while further post-training enables it to capture patterns specific to recommendation tasks~\cite{liu2025onerec, lin2025rec}. However, despite the encouraging results achieved by this line of research, the massive scale of model parameters makes it challenging to meet the stringent low-latency constraints required for online serving systems under limited computational resources. In contrast, implicit reasoning~\cite{hao2024training,chen2025reasoning, li2025implicit} does not forcibly map intermediate representations into text tokens, it completes the thinking process through internal model iterations with fewer steps, significantly reducing the computational load during inference. ReaRec~\cite{tang2025think}, LARES~\cite{liu2025lares}, LCR-SER~\cite{shi2025bridging} and OnePiece~\cite{dai2025onepiece} first explored latent reasoning in traditional sequential recommendation by autoregressively generating multiple latent status before producing the final user representation, promoting user modeling through progressive refinement of these latent status. For similar explorations in generative paradigm, STREAM-Rec~\cite{zhang2025slow} directly fills the reasoning path with the similar items, while GPR~\cite{zhang2025gpr} generates latent status from intent embeddings and refines them via diffusion. However, these methods exhibit two fundamental limitations. First, the rigid sequential separation of reasoning and generation creates imbalanced computational focus across hierarchical SID codes, degrading the quality of SID codes, leading to a mismatch with the hierarchical granularity of SIDs. Second, the latent reasoning status lacks physical semantics while their optimization suffers from absent reliable supervision, collectively undermining verifiable reasoning validity. This dual deficiency motivates our stepwise reasoning approach with hierarchical semantic supervision.

%% file: 3_preliminary.tex
\section{PRELIMINARY}

\noindent\textbf{Generative Recommendation.} Given a user's historical interaction sequence $ H = \{ \text{item}_1, \text{item}_2, \ldots, \text{item}_n \} $ where each item possesses multimodal representations. we train a tokenizer leveraging item-level multimodal features to map each item to an ordered token sequence of length $ m $, denoted as $ (\mathbf{s}_1, \mathbf{s}_2, \ldots, \mathbf{s}_m) $, termed the semantic ID (SID). The historical sequence is thereby converted into $ \hat{H} = \{ (\mathbf{s}_1, \ldots, \mathbf{s}_m), (\mathbf{s}_{m+1}, \ldots, \mathbf{s}_{2m}), \ldots, (\mathbf{s}_{(n-1)m+1}, \ldots, \mathbf{s}_{nm}) \} $, which forms a sequence of length $ n \times m $. This sequence serves as input to a Transformer-based sequence generation model, which autoregressively predicts the subsequent $ m $ tokens, corresponding to the SID of the next item the user is most likely to interact with. Formally, the predictive distribution of the next item $\hat{\text{item}}_{n+1} $ is defined as:
\begin{align}
    P\left( \hat{\text{item}}_{n+1} \mid H \right) &= P\left( \mathbf{s}_{L+1}, \mathbf{s}_{L+2}, \ldots, \mathbf{s}_{L+m} \mid \hat{H} \right) \notag \\
    &= \prod_{k=1}^m P\left( \mathbf{s}_{L+k} \mid \mathbf{s}_1, \mathbf{s}_2, \ldots, \mathbf{s}_{L+k-1} \right), 
\end{align}
where $ \mathbf{s}_{L+1}, \ldots, \mathbf{s}_{L+m} $ denote the generated tokens forming the SID of $ \hat{\text{item}}_{n+1} $.

\noindent\textbf{Problem Formulation.} In this work, we focus on enhancing the reasoning capabilities of generative recommendation. The prevailing paradigm of directly predicting the next SID code is constrained by limited computational depth, failing to fully exploit the potential of test-time scaling. We propose that before autoregressively generating the next SID code, the model should first generate stepwise thinking tokens $t$, which serve as preconditions to guide the generation of subsequent SID codes:
\begin{align}
    P\left( \hat{\text{item}}_{n+1} \mid H \right)
    &= \prod_{k=1}^m P\left( \mathbf{t}_{L+k} \mid \mathbf{s}_1, \mathbf{t}_1, \ldots, \mathbf{s}_{L+k-1} \right) \notag \\
    & \quad \quad \quad \cdot P\left( \mathbf{s}_{L+k} \mid \mathbf{s}_1, \mathbf{t}_1, \ldots, \mathbf{s}_{L+k-1},\mathbf{t}_{L+k} \right), 
\end{align}
Crucially, the design of reasoning mechanism compatible with the generative SID paradigm and their corresponding optimization objectives remains an open question.

%% file: 4_method.tex
\section{METHODOLOGY}

\subsection{Overview}
To activate deeper reasoning capabilities in generative recommendation (GR), we propose stepwise semantic-guided reasoning in latent space (S$^2$GR). First, high quality semantic mapping foundation from coarse to fine-grained levels are ensured by leveraging both the item co-occurrence relationships and codebook distribution optimization with load balance activation strategies. Next, our stepwise thinking mechanism introduces interpretable thinking tokens before each semantic ID (SID) generation step. This process is guided by coarse-grained codebook semantic supervision, which guarantees a reliable reasoning path and maintains stepwise semantic consistency.

\begin{figure*}[t]
\centering
\includegraphics[width=1.0\textwidth]{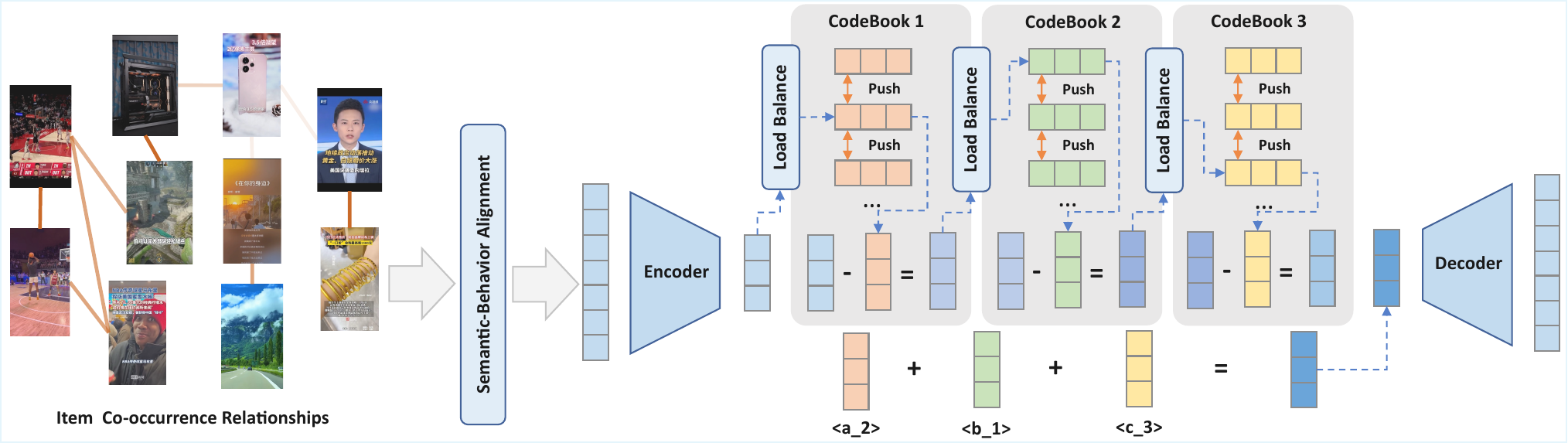} 
\caption{Framework of Collaborative and Balanced RQ-VAE. We first construct an item co-occurrence graph based on users’ historical interactions, where the color intensity indicates the frequency of co-occurrence. The original item representations are then enriched with co-occurrence information via neighborhood aggregation. During residual quantization training, we introduce a within-codebook distribution uniformity constraint to prevent codebook representation collapse. Additionally, when activating codewords, we dynamically adjust the selection probability of each codeword based on historical activation records to achieve balanced codebook usage.}
\label{fig:rq}
\end{figure*}

\subsection{Collaborative and Balanced RQ-VAE}
SIDs generated through residual quantization form the foundation of GR, where pioneering methods like RQ-VAE~\cite{rajput2023recommender} and RQ-KMeans~\cite{luo2025qarm} map high-dimensional item embeddings into discrete code sequences with hierarchical semantic via coarse-to-fine codebooks. However, these approaches exhibit certain limitations. They exclusively rely on item multimodal embeddings, disregarding valuable behavioral signals from user interaction patterns, which creates semantic misalignment between generated SIDs and downstream recommendation objectives. Moreover, performing quantization training without additional constraints can cause codebook degradation. Inadequate distribution uniformity may trigger semantic collapse in frequently used codewords, while load imbalance leaves other codewords under-activated. These issues ultimately lead to reduced codebook utilization rate (CUR) and independent coding rate (ICR). These deficiencies collectively undermine the precision of hierarchical semantic representation in SIDs. To address these challenges, we propose \underline{Co}llaborative and \underline{Ba}lanced RQ-VAE (CoBa RQ-VAE), which incorporates collaborative signals derived from item co-occurrence relationships to enhance behavioral semantics. At the same time, it improves codebook utilization by applying uniformity constraints and adaptive load-balancing mechanisms.

\subsubsection{Semantic-Behavior Alignment}
Most existing behavior alignment approaches typically pretrain collaborative filtering models to guide residual quantization~\cite{ye2025dual,li2025bbqrec,penha2025semantic,yang2025sparse}, which introduces significant computational costs. To eliminate this resource-intensive process while maintaining effective semantic-behavior alignment, we propose a lightweight approach that directly incorporates collaborative signals from raw interaction data. 

We first construct an item co-occurrence graph $\mathcal{G} = (\mathcal{V}, \mathcal{E})$, $\mathcal{U}$ denotes users's interaction records and $\mathbb{I}_{\text{co-occur}}(i,j)=1$ if item $i$ and $j$ co-occur within a specified time or size window in user's interaction history. The edge weights $w_{ij} = \sum_{u\in\mathcal{U}} \mathbb{I}_{\text{co-occur}}(i,j)$ reflect historical co-occurrence frequencies, capturing implicit behavioral relationships. This encodes essential collaborative patterns without requiring additional model training. We incorporate behavioral information through iterative propagation:
\begin{align}
    H^{(k)} &= (1 - \alpha)\widehat{A}H^{(k-1)} + \alpha X, \\
    \widehat{A} &= D^{-1/2}AD^{-1/2},
\end{align}
where the adjacency matrix $A$ encodes edge weights ($A_{ij} = w_{ij}$), while the degree matrix $D = \text{diag}(d_1,\dots,d_N)$ contains node degrees $d_i = \sum_j w_{ij}$, preserving original semantic information while integrating collaborative contexts. 

The final aligned embeddings fuse multi-hop neighborhood representations via exponentially weighted combination:
\begin{equation}
H_{\text{align}} = \sum_{k=0}^{K} \beta^k H^{(k)} \quad \text{where} \quad \beta^k = \frac{ \alpha(1-\alpha)^k }{ \sum_{i=0}^{K} \alpha(1-\alpha)^i },
\end{equation}
which intrinsically harmonize semantic content and behavioral patterns. This design eliminates training overhead, operates efficiently through sparse matrix computations, and provides high quality input for subsequent quantization by jointly preserving item semantics and behavior relationships.

\begin{figure*}[t]
\centering
\includegraphics[width=1.0\textwidth]{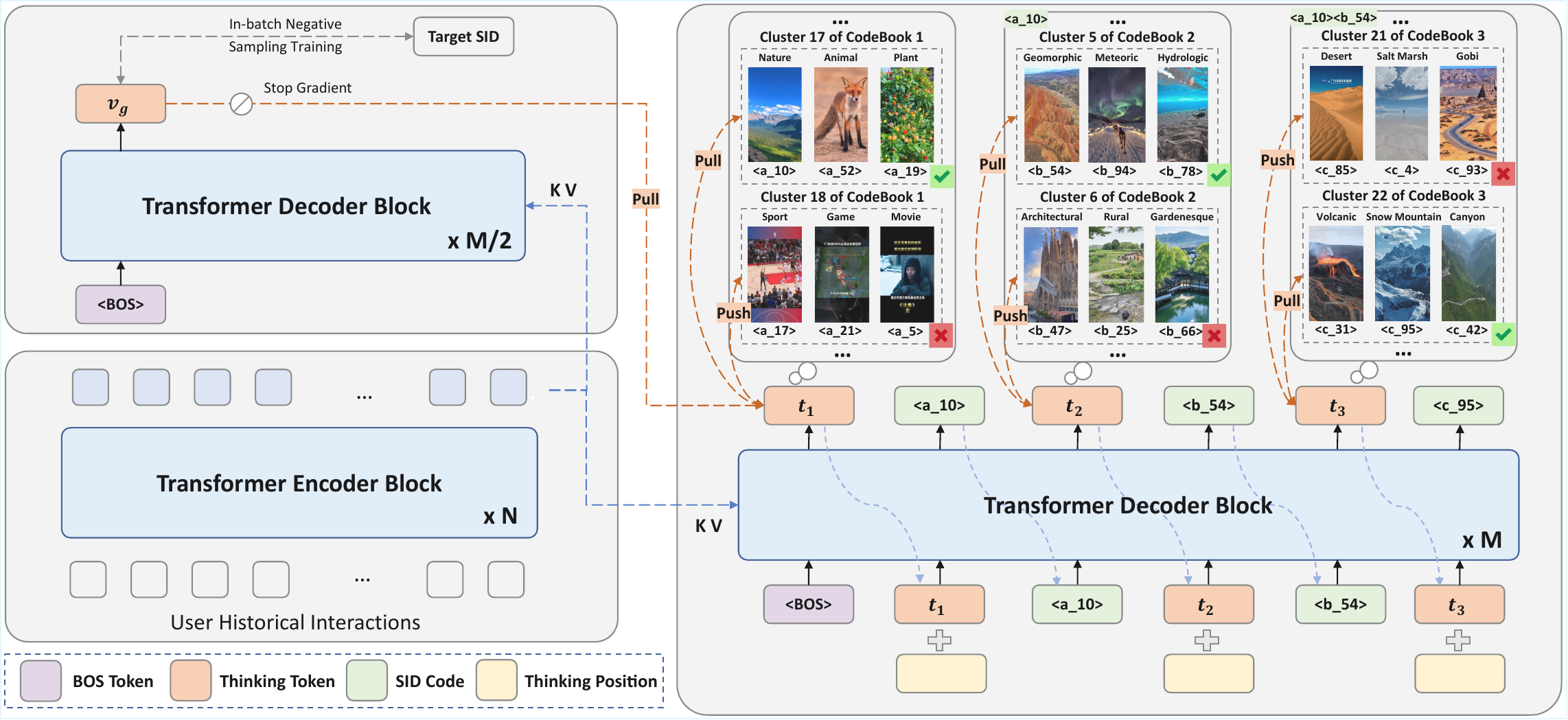} 
\caption{Framework of Stepwise Semantic-Guided Reasoning. We interleave stepwise thinking tokens within the semantic ID generation process, which represent coarse-grained category semantics, guiding the subsequent SID code to be drawn from the semantic cluster associated with that thinking token. Each thinking token is supervised by the coarse-grained semantics obtained via codebook clustering, while the initial thinking token is additionally regularized to integrate the holistic item semantics, ensuring the reliability of the reasoning path.}
\label{fig:framework}
\end{figure*}

\subsubsection{Codebook Uniform and Load Balance}

Given the behavior aligned item representation $x \in H_{\text{align}}$, we first follow the basic paradigm of RQ-VAE~\cite{rajput2023recommender}, the encoder $E$ first transforms input into a latent representation $z = E(x)$. The residual is initialized as $r_0 = z$, and at each quantization layer $l \in \{1,\dots,L\}$:
\begin{align}
\label{eq:codebook_select}
c_l &= \arg\min \{d_k\}_{k=1}^K \quad \text{where} \quad d_k = \| r_{l-1} - e_l^k \|_2^2, \\
r_l &= r_{l-1} - e_l^{c_l},
\end{align}
where $\{e_l^k\}_{k=1}^K$ denotes the codebook at layer $l$. The quantized representation $\hat{z} = \sum_{l=1}^D e_l^{c_l}$ is reconstructed via decoder $G$ as $\hat{x} = G(\hat{z})$. The reconstruction and quantization losses can be represented as:
\begin{align}
    \mathcal{L}_{\text{recon}} &= \| x - \hat{x} \|_2^2, \\
    \mathcal{L}_{\text{quant}} &= \sum_{l=1}^L \left( \| r_{l-1} - \text{sg}(e_l^{c_l}) \|_2^2 + \| e_l^{c_l} - \text{sg}(r_{l-1}) \|_2^2 \right),
\end{align}
where $\text{sg}(\cdot)$ denotes stop-gradient operation.

To prevent codebook representation collapse, where some codewords become overly similar and lose semantic discriminability, we introduce a pairwise distance-based distribution uniform loss. For codebook embeddings in layer $l$:
\begin{align}
\mathcal{D}_{ij} &= \| e_l^i - e_l^j\|_2^2, \\
\mathcal{P} &= \{(i,j) \mid \mathcal{D}_{ij} \leq \delta, i \neq j\}, \\
\mathcal{L}_{\text{uniform}} &= \log \left(1 + \sum_{(i,j) \in \mathcal{P}} \exp(-\mu \cdot \mathcal{D}_{ij}) \right),
\end{align}
where $\delta$ is a distance threshold, and $\mu$ controls penalty intensity. This formulation specifically targets clusters of similar vectors within the codebook, applying exponential penalties to encourage dispersion while ignoring distant pairs.

Meanwhile, to address skewed utilization where some codewords are frequently activated while others remain inactive, inspired by load balance mechanism in Mixture-of-Experts (MoE) models~\cite{wang2024auxiliary,liu2024deepseek,dai2024deepseekmoe}, we dynamically adjusts selection probabilities in Eq.~\eqref{eq:codebook_select} based on historical activation frequency:
\begin{align}
f_k &= \text{clip}\left( \frac{u_{\text{mean}} - u_k}{u_{\text{mean}} + \epsilon}, -\delta_{\text{max}}, \delta_{\text{max}} \right), \\
\hat{d_k} &= d_k \odot (1 - f_k),
\end{align}
where $u_k$ tracks the activation count for codeword $k$, $u_{\text{mean}}$ represents the average activation count, $\delta_{\text{max}}$ limits adjustment strength, and $\epsilon$ prevents division by zero.

\subsection{Stepwise Semantic-Guided Reasoning}

To address the problem of uneven computational focus and the lack of verifiable supervision in existing reasoning-enhanced GR methods, we propose a stepwise reasoning mechanism that naturally aligns with the hierarchical granularity of SID. Previous approaches suffer from the weakness in that the reasoning process ends before the complete multi-step generation of SID, which leads to unbalanced computational focus across different hierarchical SID code positions. Meanwhile, the absence of supervision grounded in interpretable semantics makes it difficult to ensure reliable reasoning paths. Our core insight is that semantic IDs are inherently corresponds to hierarchical semantics, and the generation process can be viewed as a multi-step decision-making process from coarse to fine granularity. Consequently, the stepwise reasoning is more compatible, where reasoning step should be interleaved with each generation step, and each reasoning step explicitly models the semantics corresponding to its level in the hierarchy and receives supervision tailored to that level. This design ensures balanced computation across all steps and provides interpretable guidance throughout the reasoning process.

\subsubsection{Stepwise Reasoning Paradigm}

Given encoder outputs $\mathbf{H}_{\text{enc}} \in \mathbb{R}^{n \times d}$ and initial [BOS] token embedding $\mathbf{e}_0 \in \mathbb{R}^d$, the model operates recursively over $l \in \{1,\dots,L\}$. At each level $l$, The decoder autoregressively generates thinking token $\mathbf{t}_{l}$ and SID code token $\mathbf{s}_{l}$:  
\begin{align}
\mathbf{\tilde{h}}_{l-1} &= (\mathbf{e}_0, \mathbf{t}_1, \mathbf{s}_1, \dots, \mathbf{t}_{l-1}, \mathbf{s}_{l-1}), \\
\mathbf{t}_l &= \text{Decoder}_{\theta}(\mathbf{\tilde{h}}_{l-1}, \mathbf{H}_{\text{enc}}) + \mathbf{p}_l, \\
\mathbf{s}_l &= \text{Decoder}_{\theta}((\mathbf{\tilde{h}}_{l-1}, \mathbf{t}_l), \mathbf{H}_{\text{enc}}),
\end{align}
where $\mathbf{p}_l \in \mathbb{R}^d$ denotes hierarchy-specific position embedding, explicitly differentiate thinking positions from SID code generation positions. This granularity-aligned reasoning structure resolves the problem of unfocused computational focus in reasoning. Through subsequent coarse-grained semantic alignment supervision, the thinking token $\mathbf{t}_l$ transitions from the latent embedding without explicit supervision into the semantic anchor that represents coarse-grained categories, thereby establishing foundational grounds for the generation of $\mathbf{s}_l$.

\subsubsection{Semantic-Guided Alignment Supervision}

To enable smoother transitions between SID generation steps, we align the thinking tokens with the coarse-grained semantics associated with the next step SID code, which primary alignment transforms the latent embeddings into semantic anchors, thereby guiding the subsequent fine-grained SID code generation. We propose coarse-grained semantic-guided alignment loss as the principal supervision signal of thinking tokens. Specifically, for thinking token $\mathbf{t}_l$, we first extract embeddings for codebook in level $l$, then perform K-means clustering to obtain centroid vectors $\mathbf{C}_l \in \mathbb{R}^{K \times d}$ representing coarse-grained semantic categories. Each next target SID code $\mathbf{s}_l^i$ is mapped to its cluster centroid $\mathbf{c}_{{\mathbf{s}_l^i}}$:
\begin{equation}
\mathcal{L}_{\mathrm{Align}}^{(l)} = -\frac{1}{B}\sum_{i=1}^{B} \log \frac{\exp(\text{Sim}(\mathbf{t}_l^{i}, \mathbf{c}_{{\mathbf{s}_l^i}}) / \tau)}{\sum_{k=1}^{K} \exp(\text{Sim}(\mathbf{t}_l^{i}, \mathbf{c}_{k}) / \tau)},
\end{equation}
where $\text{Sim}(\cdot)$ denotes measurement of cosine similarity, $\tau$ denotes temperature coefficient, and $B$ is batch size. The loss explicitly maximizes alignment between thinking tokens and their corresponding coarse-grained category centroids, ensuring that each thinking token serves as a semantic bridge before the next SID code generation. By enforcing this coarse-grained alignment, the reasoning process incorporates an intentional intermediate stage in which the model first constructs a high-level semantic representation. This representation directly corresponds to the cluster in which the ground truth SID code resides, thereby providing a semantically consistent transition from broader conceptual reasoning to precise fine-grained code prediction.

Despite the coarse-grained semantic alignment provided by $\mathcal{L}_{\mathrm{Align}}$, the initial thinking token remains critical as the foundation of the entire reasoning chain. To enhance its robustness, we embed a holistic item-level perspective into the initial thinking token, ensuring stable semantic grounding at the reasoning origin and preventing semantic shifts in the early codes. Specifically, to integrate global item semantic grounding in the initial thinking token $\mathbf{t}_1$, we introduce a lightweight decoder that generates a one-step embedding $\mathbf{v}_{g}$:
\begin{equation}
\mathbf{v}_{g} = \text{Decoder}_{\psi}(\mathbf{e}_{0}, \mathbf{H}_\text{enc}).
\end{equation}

The target embedding $\bar{\mathbf{v}}$ is obtained by averaging all hierarchical 
SID code embeddings associated with the target item, encapsulating its holistic item-level semantics. We employ in-batch negative sampling InfoNCE to align $\mathbf{v}_{g}$ with $\bar{\mathbf{v}}$:
\begin{equation}
\mathcal{L}_{\mathrm{InfoNCE}} = -\frac{1}{B}\sum_{i=1}^{B} \log \frac{\exp(\text{Sim}(\mathbf{v}_{g}^{i}, \bar{\mathbf{v}}^{i}) / \tau)}{\sum_{j=1}^{B} \exp(\text{Sim}(\mathbf{v}_{g}^{i}, \bar{\mathbf{v}}^{j}) / \tau)}.
\end{equation}

To directly inject the global semantics into the reasoning origin, we introduce a cosine similarity regularizer between $\mathbf{t}_1$ and $\mathbf{v}_{g}$:
\begin{equation}
\mathcal{L}_{\mathrm{Reg}} = 1 - \frac{1}{B} \sum_{i=1}^{B} \text{Sim}(\mathbf{t}_1, \text{sg}(\mathbf{v}_{g})),
\end{equation}
where $\text{sg}(\cdot)$ denotes stop-gradient operations. This ensures $\mathbf{t}_1$ inherits stable, item-level global semantic, mitigating initial reasoning drift and improving downstream consistency.

The overall supervision objective as follows:
\begin{align}
    \mathcal{L} &= \mathcal{L}_{\text{Rec}} + \mathcal{L}_{\text{Think}}, \\
    \mathcal{L}_{\text{Rec}} &= -\sum_{l=1}^L \text{log}~y_{\mathbf{s}_{l}}^{(l)}, \\
    \mathcal{L}_{\text{Think}} &= \sum_{l=1}^L \mathcal{L}_{\mathrm{Align}}^{(l)} + \mathcal{L}_{\mathrm{InfoNCE}} + \lambda \mathcal{L}_{\mathrm{Reg}},
\end{align}
where $\mathcal{L}_{\text{Rec}}$ denotes the supervision objective of SID tokens, $y_{\mathbf{s}_{l}}^{(l)}$ denotes the logits corresponding to SID code $\mathbf{s}_{l}$, $\mathcal{L}_{\text{Think}}$ denotes the supervision objective of thinking tokens, $\lambda$ is hyperparameter for trading-off.

%% file: 5_experiment.tex
\section{EXPERIMENTS}

\subsection{Experimental Setup}

\subsubsection{Datasets and Evaluation Metrics}

To evaluate the proposed~S$^2$GR, we conduct comprehensive experiments on datasets of various scales, as shown in Table~\ref{dataset}, including both a public dataset and an industrial dataset. For the public dataset, we use the "Beauty" subset of the Amazon Product Reviews~\cite{he2016ups}, which contains product reviews collected from May 1996 to September 2014. For the industrial dataset, we collect real interaction logs (i.e., long-play video records) from sampled active users over a consecutive 2 days period from a large-scale online short video platform. For the evaluation metrics, we adopted top-K HitRate (donated as HR@K) and Normalized Discounted Cumulative Gain (donated as NDCG@K) to assess the performance in terms of recall and ranking, respectively. The detailed calculation methods can be found in the Appendix~\ref{metrics}.

\input{tables/dataset}

\subsubsection{Baselines}

We selected a set of competitive baselines, including traditional sequential recommendation methods and emerging generative recommendation (GR) methods:

\begin{itemize}[leftmargin=0.3cm]
    \item \textbf{Caser}~\cite{tang2018personalized}: Caser models interaction sequences using convolutional neural networks to capture both point-level and union-level patterns via horizontal and vertical convolutional filters.
    \item \textbf{GRU4Rec}~\cite{hidasi2015session}: GRU4Rec adopts gated recurrent units to model interaction sequences.
    \item \textbf{SASRec}~\cite{kang2018self}: SASRec leverages a unidirectional Transformer encoder with self-attention mechanisms to model interaction sequences, focusing on left-context information to predict the next item efficiently. 
    \item \textbf{BERT4Rec}~\cite{sun2019bert4rec}: BERT4Rec utilizes a deep bidirectional Transformer encoder trained with the masked item prediction to capture contextual information from both left and right directions within interaction sequences. 
    \item \textbf{ReaRec}~\cite{tang2025think}: ReaRec is a reasoning-enhanced sequential recommendation model that progressively refines user representations through multi-step hidden states.
    \item \textbf{TIGER}~\cite{rajput2023recommender}: TIGER is a generative retrieval framework that represents items via semantic IDs (SIDs) generated by residual quantization and uses a Transformer-based sequence-to-sequence model to autoregressively predict the next item's SID.
\end{itemize}

\input{tables/exp_main}

\subsubsection{Implementation Details}

For SIDs, we adopt a three-level structure, with each level having a codebook size of 256. For Amazon Beauty, we encode item titles and descriptions using the Qwen3-Embedding-4B~\cite{qwen3embedding} model to obtain item embeddings, while for the industrial dataset, we generate embeddings from sampled video frames using a proprietary multimodal large model. Our model is based on the T5~\cite{ni2022sentence} architecture, comprising 4 encoder layers and 4 decoder layers, consistent with the configuration used in the TIGER baseline. In the codebook optimization stage of our approach, the user historical sequences used to construct the item co-occurrence relationships precede those employed for training and evaluation, in order to prevent information leakage.

\subsection{Performance Comparison}

Table~\ref{exp_main} summarizes the main results of our approach compared to state-of-the-art baselines. We observe three key findings:  
\begin{itemize}[leftmargin=0.3cm]
    \item Traditional sequential recommendation methods achieve competitive performance relative to generative methods on small-scale dataset, but exhibit significant degradation on large-scale industrial dataset. This occurs because traditional item ID-based learning struggles to generalize in massive item spaces due to sparse interactions and limited representation capacity, while generative methods leverage compressed SIDs to efficiently capture hierarchical semantic patterns.
    \item Our method consistently outperforms all baselines, with average improvements of 9.47\% and 33.75\% on public and industrial datasets respectively. This superiority stems from the stepwise reasoning mechanism with reliable supervision which ensures the balanced computational focus while providing effective guidance and smooth transitions for SID code generation. Meanwhile, the integration of collaborative information with a balanced optimized codebook further enhances overall performance.
    \item Performance gains are more pronounced on large-scale dataset compared to small-scale dataset. The longer interaction sequences in industrial dataset better activate the reasoning capability, allowing hierarchical thinking tokens to leverage coarse-grained semantic context for more accurate fine-grained predictions.
\end{itemize}

\subsection{Ablation Study}

We analyze three variants to validate their contributions:
\begin{itemize}[leftmargin=0.3cm]
    \item \textbf{S$^2$GR}~(\textit{w/o} CoBa RQ-VAE): We replace CoBa RQ-VAE with vanilla RQ-VAE as the SID tokenization method, discarding collaborative information, distribution uniform loss, and load balance.
    \item \textbf{S$^2$GR}~(\textit{w/o} Reason): We eliminate the stepwise thinking tokens and associated supervision mechanisms, thereby generating SID without the enhancement of latent reasoning.
    \item \textbf{S$^2$GR}~(\textit{w/o} $\mathcal{L}_{\mathrm{Think}}$): We retain the stepwise thinking tokens but remove the associated supervision mechanisms.
\end{itemize}

Overall, significant performance degradation occurs across all variants, confirming each component's necessity. Specifically,                      
\textbf{(1)} For \textbf{S$^2$GR}~(\textit{w/o} CoBa RQ-VAE), as shown in Table~\ref{exp_rq}, our optimized codebook achieves significant improvement in codebook utilization rate (CUR) and independent coding rate (ICR). Removing this component weakens the hierarchical semantic foundation, causing semantic conflicts in codewords, which leads to misaligned reasoning paths during SID generation. 
\textbf{(2)} For \textbf{S$^2$GR}~(\textit{w/o} Reason), removing the stepwise reasoning tokens disables the stepwise semantic alignment process, causing the model to degrade into direct SID prediction without coarse-grained thinking process, which reduces the overall quality of the generated SIDs. 
\textbf{(3)} For \textbf{S$^2$GR}~(\textit{w/o} $\mathcal{L}_{\mathrm{Think}}$), removing the supervision mechanism for the thinking tokens causes them to degenerate into latent states without explicit meaning. While simply extending the thinking process to increase computational depth can be beneficial for enhancing reasoning performance, in the absence of supervision, the guidance provided by the thinking tokens for SID generation is weaker, and the resulting reasoning process may not be entirely reliable.

\input{tables/exp_rq}

\input{tables/exp_ab}

\subsection{Online Testing}

To verify the effectiveness of~S$^2$GR~in a real industrial environment, we deployed it on a large-scale short video platform for an online A/B test. S$^2$GR was used as the experimental group model, while TIGER~\cite{rajput2023recommender} served as the control group model, with each group assigned 5.25\% of users, and the experiment lasted for 7 days, to ensure the reliability of the statistical results. As shown in Table~\ref{exp_ab_test},~S$^2$GR~significantly increased users’ app usage duration as well as video viewing frequency, demonstrating the effectiveness of our method in large-scale real-world scenarios.

\subsection{Further Analysis}

\subsubsection{Analysis of hyperparameter sensitivity}
\begin{figure}[t]
\centering
\includegraphics[width=\linewidth]{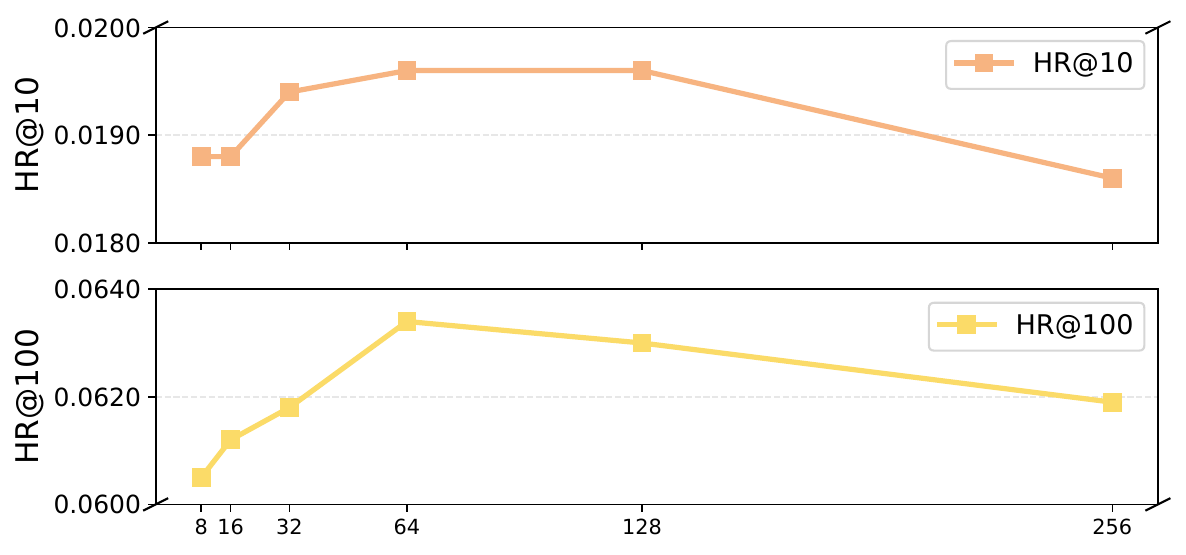} 
\caption{The performance variation across different cluster number of coarse-grained semantic guidance.}
\label{fig:hyper}
\end{figure}

To investigate the sensitivity of hyperparameters, we examine the primary hyperparameter in our method, namely the number of clusters used to generate step-wise semantic guidance from the codebook. As shown in Figure~\ref{fig:hyper}, the performance initially improves as the number of clusters increases, which can be attributed to the coarse-grained semantic guidance for the thinking tokens becoming relatively more fine-grained. However, as the number continues to grow, the performance starts to decline, which is because overly fine-grained semantic guidance leads the thinking tokens to "overthink", thereby failing to function as a transitional stage and to provide effective guidance for the generation of subsequent sid codes.

\subsubsection{Analysis of user interaction sequences of different lengths}
\begin{figure}[t]
\centering
\includegraphics[width=\linewidth]{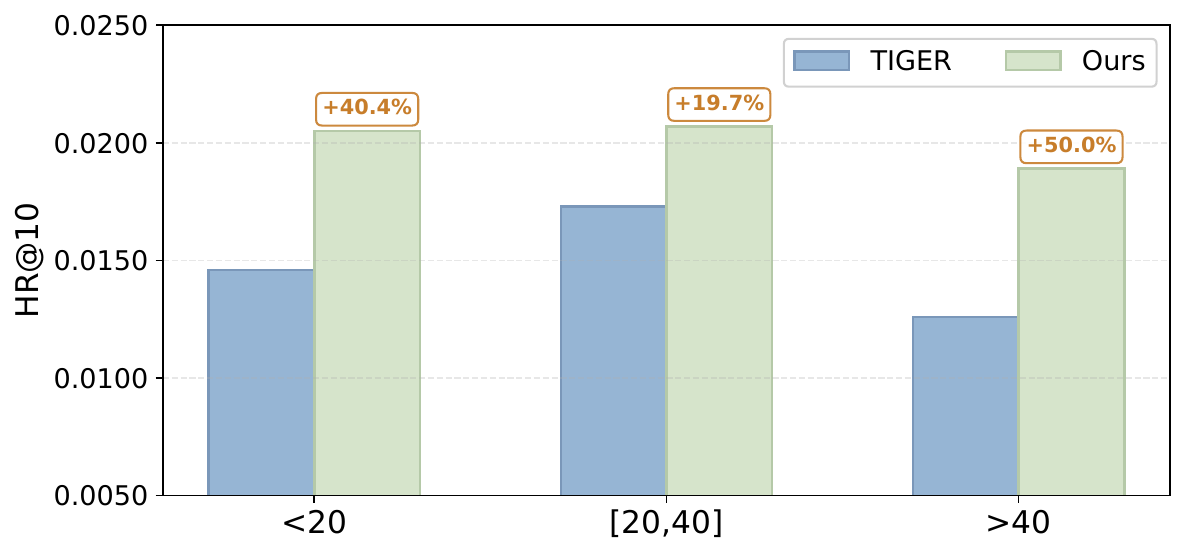} 
\caption{The performance comparison under different user interaction sequence lengths on industrial dataset.}
\label{fig:seqlen}
\end{figure}

We evaluate the impact of~S$^2$GR~across different user interaction sequence lengths on industrial dataset, as shown in Figure~\ref{fig:seqlen}. We observe that the~S$^2$GR~achieves higher relative improvements when applied to short and long interaction sequences, whereas the gains are more modest for sequences of moderate length. For short sequences, the limited amount of contextual information often leads to lower performance without reasoning process, and the incorporation of reasoning enables the model to infer latent user interests, yielding noticeable gains. For long sequences, the abundance of interaction data introduces complexity and potential noise, and reasoning helps to distill salient patterns from this rich but intricate information, further revealing user preferences. In contrast, for moderate-length sequences, the available context is already sufficient for vanilla method to perform effectively, so the additional benefit from reasoning is comparatively smaller.

\subsubsection{Analysis of different model size}
\begin{figure}[t]
\centering
\includegraphics[width=\linewidth]{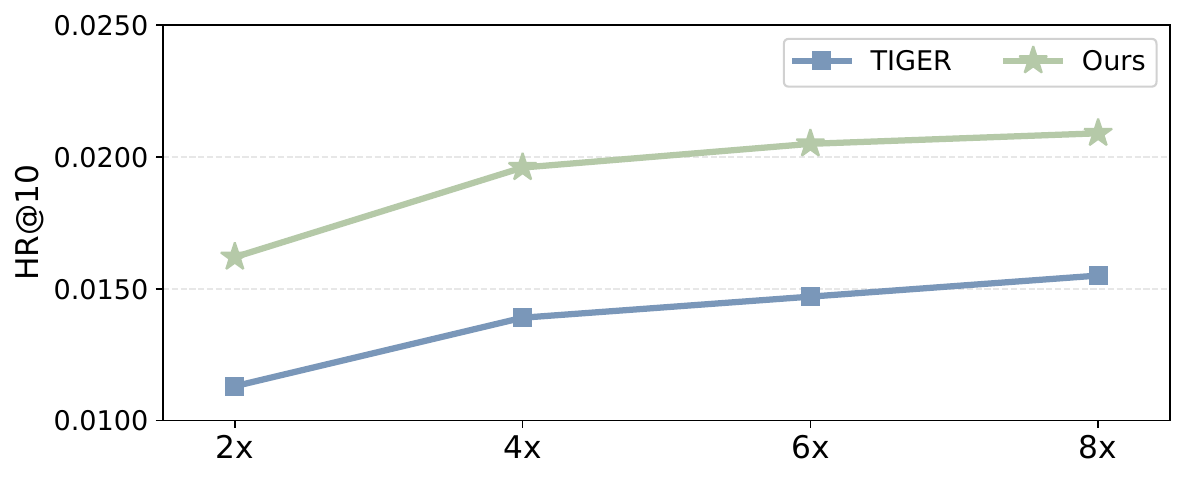} 
\caption{The performance comparison under different model parameters on industrial dataset.}
\label{fig:para}
\end{figure}

To investigate the effect of model size, we uniformly adjusted the number of encoder and decoder layers for both TIGER and our method. As shown in Figure~\ref{fig:para}, the experimental results demonstrate that, as increasing model size, both methods demonstrate performance improvement. Nevertheless, our method consistently maintains substantial and stable advantages over the baseline across all scales, indicating that the observed benefits are not confined to specific configurations but remain consistent as the model capacity grows.

%% file: tables/dataset.tex
\setlength{\equalcolwidth}{0.145\textwidth} 

\begin{table}[t]
\renewcommand{\arraystretch}{1.1}
  \centering
  \caption{The statistics of experimental datasets.}  
    \resizebox{1.0\linewidth}{!}{
    \setlength{\tabcolsep}{2.1pt}
    \begin{tabular}{@{}c*{3}{>{\centering\arraybackslash}p{\equalcolwidth}}@{}}
    \toprule

    \textbf{Dataset} & \textbf{Amazon Beauty} & \textbf{Industrial Dataset} \\
    \midrule
    
    \#User & 22363 & 37500\\
    \#Item & 12101 & 1086872\\
    \#Inter & 198502 & 3778562\\
    \#Avg.Inter. / User & 8.88 & 100.76\\
    \#Avg.Inter. / Item & 16.40 & 3.48\\
    Sparisty & 99.92\% & 99.99\%\\
    
    \bottomrule
    \end{tabular}}
  \label{dataset}%

\end{table}%

%% file: tables/exp_main.tex
\setlength{\equalcolwidth}{0.085\textwidth} 

\begin{table*}[t]
\renewcommand{\arraystretch}{1.1}
  \centering
  \caption{Overall performance on public dataset and industrial-scale dataset. The best performance is highlighted in bold while the second best performance is underlined.}  
\resizebox{1.0\linewidth}{!}{
    \setlength{\tabcolsep}{2.1pt}
    \begin{tabular}{@{}cc*{8}{>{\centering\arraybackslash}p{\equalcolwidth}}@{}}%
    \toprule
    \multirow{2}{*}{\textbf{Method}} & \multicolumn{4}{c}{\textbf{Amazon Beauty}$^1$} & \multicolumn{4}{c}{\textbf{Industrial Dataset}} \\ 
    \cmidrule(lr){2-5} \cmidrule(lr){6-9} & HR@5   & HR@10  & NDCG@5 & NDCG@10 & HR@10   & HR@100 & NDCG@10   & NDCG@100 \\ 
    
    \midrule

        \textbf{Caser} & 0.0314 & 0.0514 & 0.0186 & 0.0251 & 0.0038 & 0.0178 &  0.0020 & 0.0046 \\ 
        \textbf{GRU4Rec} & 0.0384 & 0.0634 & 0.0233 & 0.0314 & 0.0056 & 0.0251 & 0.0027 & 0.0065\\ 
        \textbf{SASRec} & 0.0434 & 0.0688 & 0.0259 & 0.0340 & 0.0088 & 0.0378 & 0.0046 & 0.0102  \\ 
        \textbf{BERT4Rec} & 0.0394 & 0.0627 & 0.0255 & 0.0329 & 0.0067 & 0.0280 & 0.0032 & 0.0072 \\ 

        \textbf{ReaRec} & \underline{0.0460} & 0.0723 & 0.0284 & 0.0368 & 0.0102 & 0.0354 & 0.0059 & 0.0105 \\
        
        \textbf{TIGER} & 0.0443 & \underline{0.0728} & \underline{0.0299} & \underline{0.0390} & \underline{0.0139} & \underline{0.0560} & \underline{0.0072} & \underline{0.0154} \\ 
    
    \midrule
    
        \textbf{S$^2$GR} & \textbf{0.0495} & \textbf{0.0785} & \textbf{0.0340} & \textbf{0.0424} & \textbf{0.0196} & \textbf{0.0634} & \textbf{0.0111} & \textbf{0.0195} \\

        Improv. & \(\uparrow\)~7.61\% & \(\uparrow\)~7.83\% & \(\uparrow\)~13.71\% & \(\uparrow\)~8.72\% & \(\uparrow\)~41.01\% & \(\uparrow\)~13.21\% & \(\uparrow\)~54.17\% & \(\uparrow\)~26.62\% \\

    \midrule

        \makecell[c]{\textbf{S$^2$GR} (\textit{w/o} CoBa RQ-VAE)} 
        & \multicolumn{1}{c}{0.0479} 
        & \multicolumn{1}{c}{0.0764} 
        & \multicolumn{1}{c}{0.0326} 
        & \multicolumn{1}{c}{0.0414} 
        & \multicolumn{1}{c}{0.0155} 
        & \multicolumn{1}{c}{0.0576} 
        & \multicolumn{1}{c}{0.0081} 
        & \multicolumn{1}{c}{0.0161} \\

        \makecell[c]{\textbf{S$^2$GR} (\textit{w/o} Reason)} 
        & \multicolumn{1}{c}{0.0471} 
        & \multicolumn{1}{c}{0.0736} 
        & \multicolumn{1}{c}{0.0321} 
        & \multicolumn{1}{c}{0.0406} 
        & \multicolumn{1}{c}{0.0167} 
        & \multicolumn{1}{c}{0.0579} 
        & \multicolumn{1}{c}{0.0093} 
        & \multicolumn{1}{c}{0.0172} \\

        \makecell[c]{\textbf{S$^2$GR} (\textit{w/o} $\mathcal{L}_{\mathrm{Think}}$)} 
        & \multicolumn{1}{c}{0.0481} 
        & \multicolumn{1}{c}{0.0740} 
        & \multicolumn{1}{c}{0.0320} 
        & \multicolumn{1}{c}{0.0402} 
        & \multicolumn{1}{c}{0.0178} 
        & \multicolumn{1}{c}{0.0589} 
        & \multicolumn{1}{c}{0.0100} 
        & \multicolumn{1}{c}{0.0180} \\

    \bottomrule
    \end{tabular}}
  \label{exp_main}%

\end{table*}%

\def\thefootnote{$^1$}\footnotetext{To ensure the reliability of the evaluation results, we filtered out user interaction sequences with a length of less than 10 during testing.}

%% file: tables/exp_rq.tex
\setlength{\equalcolwidth}{0.07\textwidth} 

\begin{table}[t]
\renewcommand{\arraystretch}{1.1}
  \centering
  \caption{Performance comparison of different SID tokenization methods.}  
    \resizebox{1.0\linewidth}{!}{
    \setlength{\tabcolsep}{2.1pt}
    \begin{tabular}{@{}c*{4}{>{\centering\arraybackslash}p{\equalcolwidth}}@{}}
    \toprule

    \multirow{2}{*}{\textbf{Method}} & \multicolumn{2}{c}{\textbf{Amazon Beauty}} & \multicolumn{2}{c}{\textbf{Industrial Dataset}} \\ 
    \cmidrule(lr){2-3} \cmidrule(lr){4-5} & CUR & ICR & CUR & ICR \\ 

    \midrule
    
    \textbf{RQ-Kmeans} & 0.062\% & 86.65\% & 4.11\% & 63.56\%  \\

    \textbf{RQ-VAE} & 0.069\% &  95.86\% & 4.61\% & 71.23\%  \\

    \textbf{CoBa RQ-VAE} & \textbf{0.072\%} & \textbf{99.30\%} & \textbf{4.95\%} & \textbf{76.25\%} \\
    
    \bottomrule
    \end{tabular}}
  \label{exp_rq}%

\end{table}%

%% file: tables/exp_ab.tex
\setlength{\equalcolwidth}{0.15\textwidth} 

\begin{table*}[t]
\renewcommand{\arraystretch}{1.1}
  \centering
  \caption{The results of the online A/B test. Confidence intervals (CI) are calculated with 0.05 significance level.}  
    \resizebox{1.0\linewidth}{!}{
    \setlength{\tabcolsep}{2.1pt}

    \begin{tabular}{@{}>{\centering\arraybackslash}p{\equalcolwidth}*{6}{>{\centering\arraybackslash}p{\equalcolwidth}}@{}}
        \toprule
        \multicolumn{2}{c}{\textbf{Total App Usage Time}} & \multicolumn{2}{c}{\textbf{App Usage Time per User}} & \multicolumn{2}{c}{\textbf{Total Video View}} \\ 
        \cmidrule(lr){1-2} \cmidrule(lr){3-4} \cmidrule(lr){5-6} 
        \textbf{Improv.} & \textbf{CI} & \textbf{Improv.} & \textbf{CI} & \textbf{Improv.} & \textbf{CI} \\ 
        \midrule
        +0.092\% & \text{[+0.03\%, +0.16\%]} & +0.088\% & \text{[+0.03\%, +0.15\%]} & +0.091\% & \text{[+0.01\%, +0.17\%]}\\ 
        \bottomrule
    \end{tabular}
        
    }

  \label{exp_ab_test}%

\end{table*}%

%% file: 6_conclusion.tex
\section{CONCLUSION}

In this paper, we propose~S$^2$GR, a novel generative recommendation (GR) framework that addresses limitations in reasoning-enhanced GR through stepwise latent reasoning with hierarchical semantic supervision. Our method first establishes semantically robust foundations by integrating item co-occurrence relationships, load balancing, and uniformity objectives, ensuring coarse-to-fine semantic granularity. The core innovation inserts interpretable thinking tokens before each semantic ID (SID) generation step supervised via contrastive alignment with ground-truth codebook clusters. This guarantees balanced computational focus across hierarchical SID codes, and reliable reasoning path. Extensive experiments demonstrate substantial improvements over state-of-the-art baselines on public dataset and industrial dataset. Online A/B tests on a large-scale short-video platform further confirm significant gains in user engagement metrics validating real-world efficacy.

%% file: 7_appendix.tex
\setcounter{figure}{0}
\setcounter{table}{0}
\setcounter{equation}{0}
\setcounter{section}{0}
\renewcommand\thesection{\Alph{section}}
\renewcommand\thefigure{\thesection.\arabic{figure}}
\renewcommand\thetable{\thesection.\arabic{table}}
\renewcommand\theequation{\thesection.\arabic{equation}}

\section{Evaluation Metrics}
\label{metrics}
We used two common evaluation metrics: HitRate@K (HR@K) and Normalized Discounted Cumulative Gain@K (NDCG@K), which measure retrieval and ranking performance, respectively.

\noindent\textbf{HR@K} indicates the proportion of ground-truth item appears within the top-$K$ predictions:
\begin{equation}
\mathrm{HR@K} = \frac{1}{N} \sum_{i=1}^{N} \mathbb{I} \left( \mathrm{rank}_i(g_i) \leq K \right),
\end{equation}
where $N$ is the number of test samples, $g_i$ is the ground-truth item for sample $i$, $\mathrm{rank}_i(g_i)$ is its predicted position, and $\mathbb{I}(\cdot)$ equals 1 if the condition is true and 0 otherwise.

\noindent\textbf{NDCG@K} considers the rank position of the ground truth item:
\begin{equation}
\mathrm{DCG@K} = \sum_{p=1}^{K} \frac{\mathrm{rel}_p}{\log_2(p+1)}, \quad
\mathrm{NDCG@K} = \frac{\mathrm{DCG@K}}{\mathrm{IDCG@K}},
\end{equation}
where $\mathrm{rel}_p$ is the relevance (1 if the item at position $p$ is the ground truth, else 0), and $\mathrm{IDCG@K}$ is the DCG@K for an ideal ranking.

\section{Supplementary Experiments}
\label{}

\subsection{Analysis of training and inference efficiency}
We compare the training and inference times of same parameter-scale models on industrial dataseton with 2×A100 GPUs. As shown in Table~\ref{exp_effi}, compared with TIGER, our method requires moderately longer training time under the same number of epochs, owing to the expansion of thinking steps and the incorporation of additional supervision. For inference, the increase in latency is relatively small. In our implementation, each decoding beam produces the thinking token and its subsequent SID code in a single continuous step, without separately expanding the beam for the thinking token, thereby avoiding multiple computational cost.

\input{tables/exp_effi}

%% file: tables/exp_effi.tex
\begin{table}[h]
\renewcommand{\arraystretch}{1.1}
  \centering
  \caption{Comparison of training and inference efficiency.}  
    \resizebox{1.0\linewidth}{!}{
    \setlength{\tabcolsep}{2.1pt}
    \begin{tabular}{@{} >{\centering\arraybackslash}p{2.5cm} >{\centering\arraybackslash}p{2.5cm} >{\centering\arraybackslash}p{2.5cm} @{}}
    \toprule

    \textbf{Method} & Training & Inference \\ 

    \midrule
    
    \textbf{TIGER} & \textasciitilde 45h & \textasciitilde 0.88h\\

    \textbf{S$^2$GR} & \textasciitilde 59h & \textasciitilde 0.90h\\

    \bottomrule
    \end{tabular}}
  \label{exp_effi}%

\end{table}%